\documentclass[letterpaper, twocolumn]{jpsj3}
\usepackage{txfonts}

\usepackage{siunitx}
\usepackage{textcomp}

\usepackage{algpseudocode}
\usepackage{algorithm}

\usepackage{here}

\usepackage{tabularx}

\usepackage{graphicx}
\usepackage{float}
\usepackage{textcomp}
\usepackage{amsmath}
\usepackage{listings}
\usepackage{url}

\usepackage{tikz}
\usetikzlibrary{trees}

\usepackage[whole]{bxcjkjatype}

\title{Hybrid Algorithm of Linear Programming Relaxation and Quantum Annealing}

\author{Taisei Takabayashi$^1$ and Masayuki Ohzeki$^{1,2,3,4}$}
\inst{Graduate School of Information Sciences, Tohoku University, Miyagi 980-8564, Japan$^1$ \\
Department of Physics, Tokyo Institute of Technology, Tokyo, 152-8551, Japan$^2$ \\
International Research Frontier Initiative, Tokyo Institute of Technology, Tokyo, 108-0023, Japan$^3$ \\
Sigma-i Co., Ltd., Tokyo, 108-0075, Japan$^4$
} 

\abst{
The demand for classical-quantum hybrid algorithms to solve large-scale combinatorial optimization problems using quantum annealing (QA) has increased. 
One approach involves obtaining an approximate solution using classical algorithms and refining it using QA. 
In previous studies, such variables were determined using molecular dynamics (MD) as a continuous optimization method. 
We propose a method that uses the simple continuous relaxation technique called linear programming (LP) relaxation.
Our method demonstrated superiority through comparative experiments with the minimum vertex cover problem versus the previous MD-based approach. Furthermore, the hybrid approach of LP relaxation and simulated annealing showed advantages in accuracy and speed compared to solving with simulated annealing alone.}

\begin{document}
\maketitle
\renewcommand{\algorithmicrequire}{\textbf{Input:}}
\renewcommand{\algorithmicensure}{\textbf{Output:}}

\section{Introduction.}

Quantum annealing (QA) is a generic algorithm that uses quantum fluctuations to solve combinatorial optimization problems\cite{kadowaki_quantum_1998, ohzeki_quantum_2011}. 
In recent years, D-Wave Systems has developed a device known as the D-Wave quantum annealer, which implements QA and enables commercial use, leading to significant attention to QA as a novel computing approach\cite{johnson_quantum_2011}.

However, as a physical device, the D-Wave quantum annealer is subject to external noise and other factors, making QA performance less than ideal. Consequently, the output may not always be the optimal solution. 
Nonetheless, as mentioned earlier, many combinatorial optimization problems can be formulated as exploring the ground state of the Ising model, making the application of QA using the D-Wave quantum annealer the subject of extensive research across various industries\cite{lucas_ising_2014}.

Various applications of QA are proposed in many fields, such as finance \cite{rosenberg2016solving, orus2019forecasting, venturelli2019reverse}, traffic flow \cite{neukart2017traffic,hussain2020optimal,inoue2021traffic}, manufacturing \cite{venturelli2016quantum, Yonaga2022, Haba2022}, logistics \cite{feld2019hybrid,ding2021implementation}, material experiments\cite{Tanaka2023}, marketing \cite{nishimura2019item}, and decoding problems \cite{IdeMaximumLikelihoodChannel2020, Arai2021code}. The benchmark test of model-based Bayesian optimization using QA was also performed\cite{Koshikawa2021}. 
A comparative study of QA with other combinatorial optimization solvers was conducted\cite{Oshiyama2022}. The quantum effects when there are multiple optimal solutions have been investigated\cite{Yamamoto2020, Maruyama2021}. Furthermore, applications of QA for machine learning as an optimization algorithm have also been reported \cite{neven2012qboost,khoshaman2018quantum,o2018nonnegative, Amin2018, Kumar2018, Arai2021, Sato2021, Urushibata2022, hasegawa2023}.

However, due to implementation constraints on the number of qubits and the influence of external noise, the current D-Wave quantum annealer faces challenges in solving large-scale combinatorial optimization problems. 
Yet, many real-world problems exceed the size that the D-Wave quantum annealer can handle. Hybrid methods combining QA with classical algorithms have been proposed to address such issues \cite{Okada2019, chancellor_modernizing_2017, Ajagekar_2020, Atobe_2020}. Hybrid schemes for solving constrained optimization problems with QA have also been proposed \cite{Yonaga2020, Ohzeki2020, hirama2023efficient}. D-Wave Systems has also released the Leap Hybrid Solver as a hybrid solution using QA and classical heuristics\cite{leap_hybrid}.

A hybrid QA approach was proposed in prior research, utilizing molecular dynamics (MD) as a classical preprocessing method\cite{Irie_HQA_2021}. 
We refer to this method as MD+QA. 
Here MD involves repeatedly solving the Hamiltonian equations to compute the time evolution of the classical Hamiltonian. 
In the previous study, the classical Hamiltonian was designed to mimic the time-dependent Hamiltonian of QA, enabling MD to find the ground state of the Ising model approximately.

In MD+QA, the focus is on the trajectory of continuous variables obtained through MD. 
These continuous variables take real values, oscillating around zero during the initial stages of MD, alternating between positive and negative values. 
As time evolution progresses, many variables settle into positive or negative values. 
Those variables that have stabilized into positive or negative values during the time evolution are considered to have their solutions determined at the MD preprocessing stage. 
Their signs are fixed as the final Ising variable states ($+1$ or $-1$).

However, some continuous variables continue to oscillate and do not converge to positive or negative values by the end of the time evolution. 
In MD+QA, such unstable variables are solved using QA. 
Since only a subset of variables exhibits instability and the majority converges to positive or negative values, the number of variables to be solved using QA is significantly reduced. 
The unstable variables obtained through MD can be seen as an extraction of "difficult" variables among all variables.
Thus, MD+QA extracts a subproblem composed of "difficult" variables through preprocessing with MD, subsequently solving them using QA.

In this study, we propose a method that combines QA with linear programming  (LP) relaxation.
This simple continuous relaxation technique converts integer programming problems into ones with real-valued variables, which are then solved as linear programming problems. 
The LP relaxation has applications in various fields, including exact methods for combinatorial optimization problems, such as branch and bound. 
Our research focuses on the minimum vertex cover problem (min-VC) as a representative combinatorial problem that can be solved using LP relaxation. When solving min-VC with LP relaxation, the solutions are known to take values of either $\{0, 1/2, 1\}$ \cite{nemhauser1974properties}.

We leverage this property and develop a novel method called LP+QA, where we use LP relaxation to identify variables that converge to $1/2$, and then solve only those variables using QA.

This paper is organized into six sections. Sect. 1, which corresponds to this section, introduces this paper. Sect. 2 describes QA and MD+QA. In Sect. 3, we introduce min-VC as the problem setting for our study. We then present the properties of the min-VC when solved by LP relaxation. We propose a new hybrid algorithm that utilizes the properties of the min-VC when solved with the LP relaxation problem in Sect. 4. Sect. 5 describes
experiments to assess the performance of our method. 
In Sect. 6, we summarize our study and discuss our prospects for the future.

\section{Background}
\subsection{Quantum Annealing}
Quantum annealing (QA) is a meta-heuristics method for solving combinatorial optimization problems\cite{Kadowaki1998}. An optimization problem is the problem of minimizing or maximizing an objective function under given constraints. 
Combinatorial optimization problems involve discrete values for the variables in the objective function and constraints.
It is known that many real-world combinatorial optimization problems can be attributed to the problem of searching for the ground state of the following Ising model:
\begin{equation}
    \mathcal{H}_{\rm Ising}(\Vec{s})=\frac{1}{2} \sum_{i\ne j}^{N}J_{ij}s_i s_j +\sum_{i=1}^{N}h_i s_i,
    \label{Ising model}
\end{equation}
where the variable $s_i \in \{+1, -1\}$ denotes the $i$-th spin. The two values correspond to the spin orientations (upward and downward). $N$ denotes the total number of spins and $J_{ij}$ represents the interaction between adjacent spins. 
In addition, $h_i$ is the local longitudinal magnetic field acting on the $i$-th spin. 
The ground state of the Ising model refers to the combination of spins $s_i \in \{ \pm 1\} \ (i=1,\cdots, N)$ that minimizes Eq.\eqref{Ising model}.

QA solves the combinatorial optimization problem by finding the ground state of the Ising model. 
Specifically, the ground state of the Ising model is searched by evolving the following quantum Hamiltonian $\mathcal{H}_{\rm QA}$ in time:
\begin{equation}
\begin{split}
&\mathcal{H}_{\rm QA}(\sigma, \tau) = \\
    &\quad A(\tau)\left[ - \sum_{i=1}^{N}\sigma_i^x \right] + B(\tau) \left[ \frac{1}{2}\sum_{i\ne j}^{N}J_{ij} \sigma_i^z \sigma_j^z \sum_{i=1}^{N}h_i \sigma_i^z\right],
\label{QA Hamiltonian}
\end{split}
\end{equation}
where $\sigma_i^x$ and $\sigma_i^z$ denote the $x$ and $z$ components of the Pauli matrix, respectively.
In addition, $\tau$ is a function that depends on time $t$, and its range is $[0, 1]$. 

The annealing schedule $A(\tau)$ and $B(\tau)$ is chosen such that at the beginning of the time evolution ($\tau=0$), $A(0)\gg B(0)$, and at the end of the time evolution ($\tau=1$), $A(1) \ll B(1)$.

The first term in Eq. \eqref{QA Hamiltonian} stands for the transverse field term, and the second term corresponds to the Hamiltonian in Eq. \eqref{Ising model}. At the starting point $\tau=0$, the transverse field term in the first term is dominant. 
This implies that the ground state is given by a superposition state of all possible combinations of the $N$ spins. 
As the time evolution progresses towards the end at $\tau = 1$, the Ising model's Hamiltonian in the second term becomes dominant. 
With sufficiently long time evolution, its ground state can be obtained\cite{suzuki_residual_2005}.

\subsection{Hybrid Quantum Annealing via Molecular Dynamics}

Molecular Dynamics (MD) is a technique for tracking the motion of numerous particles by numerically solving their equations of motion. 
The equations of motion are given by the Hamiltonian equations of motion for a classical Hamiltonian represented in terms of general coordinates $\varphi_i$ and general momenta $p_i$, where the subscript $i$ denotes the index of each particle. 
The classical Hamiltonian $\mathcal{H}$ is a function that represents the energy of the entire system\cite{Haile_J_M_1992-05}. 
By designing an appropriate classical Hamiltonian $\mathcal{H}$, one can perform energy minimization by repeatedly solving the Hamiltonian equations and calculating the time evolution of the particles.

In prior research for this paper, a classical Hamiltonian $\mathcal{H}$ was designed to explore the ground state of the Ising model, aiming to obtain approximate solutions to combinatorial optimization problems using MD. 
The Hamiltonian is formulated as follows \cite{Irie_HQA_2021}:
\begin{equation}
\begin{split}
    \mathcal{H}_{\rm{MD}}(\varphi,p ; \tau)= 
    &\alpha(\tau)\sum_{i=1}^{N}\left( \frac{p_i^{2}}{2}+V(\varphi_i) \right) + \\
    &\quad \beta(\tau)\left[ \frac{1}{2}\sum_{i \ne j}^{N}J_{ij}\varphi_i \varphi_j + \sum_{i=1}^{N}h_i|\varphi_i|\varphi_i \right].
    \label{HMD}
\end{split}
\end{equation}
The variable $\varphi_i$ corresponds to the spin variable in the Ising model. 
However, unlike the Ising spin variables with only two states $\{+1, -1\}$, $\varphi_i$ is a quantity that includes its magnitude $|\varphi_i|$.

The time evolution in MD is described by the dimensionless parameter $\tau=t/t_{f}\in[0,1]$, where $t\in [0,t_{f}]$ corresponds to the actual physical time. 
The potential term $V(\varphi)$ is a concave function given by $V(\varphi)=\varphi^{M}$, where $M=4,6,8,\cdots$. 
This form of  potential serves the purpose of preventing the divergence of the continuous-relaxation Ising model.

In addition, the scheduling functions $\alpha(\tau)$ and $\beta(\tau)$ are chosen similarly to those used in QA. 
At the beginning, $\alpha(0)\gg \beta(0)$, and at the end, $\alpha(1)\ll \beta(1)$. These choices of scheduling functions help control the annealing process effectively.
\begin{equation}
\alpha(\tau)=\alpha_f (\tau+\rho_1(1-\tau)+\rho_2 \tau(\tau-1)),
    \label{schedule_alpha}
\end{equation}
\begin{equation}
\beta(\tau)=\beta_f (\tau+\kappa_1(1-\tau)+\kappa_2 \tau(\tau-1)).
    \label{schedule_beta}
\end{equation}

To calculate the time evolution of the flux variables, we solve the equations of motion derived from the classical Hamiltonian $\mathcal{H}_{\rm{MD}}$ for the system. 
The equations of motion are typically derived using the principles of classical mechanics, especially from the following Hamiltonian equations:
\begin{equation}
    g \frac{d\varphi_i}{d\tau}=\frac{\partial \mathcal{H}_{\rm{MD}}(\varphi,p ; \tau)}{\partial p_i},\quad g \frac{d p_i}{d\tau}=-\frac{\partial \mathcal{H}_{\rm{MD}}(\varphi,p ; \tau)}{\partial \varphi_i},
    \label{Time Evolution}
\end{equation}
where $\tau=(t/{t_{f}})\equiv g t$. When $g\to 0$, the motion of the flux variables becomes adiabatic. 
The Eq. \eqref{Time Evolution} is solved numerically using the leap-frog method. 
The leap-frog method is a numerical integration technique frequently used in classical mechanics calculations.
As for the initial conditions, we set $\varphi_i(\tau=0)=0$, and $p_i(\tau=0)$ is randomly chosen from either $+1$ or $-1$.

Fig. \ref{fig:flux_orbit} below illustrates the trajectories of flux variables obtained when the designed classical Hamiltonian $\mathcal{H}_{\rm{MD}}$ is evolved using MD.

In this study, the parameters $J_{ij}$ and $h_i$ of the Ising model were randomly selected from the interval $-1 \le J_{ij} \le +1$ and $-2 \le h_i \le +2$, respectively. 
The figure represents the trajectories of all flux variables $\{\varphi_i\}\ (i=1,\cdots,10,000)$ when the total number of MD steps is set to $\delta \tau=1\ / \ 50,000$.

\begin{figure}[tbh]
\centering
\includegraphics[scale=0.45]{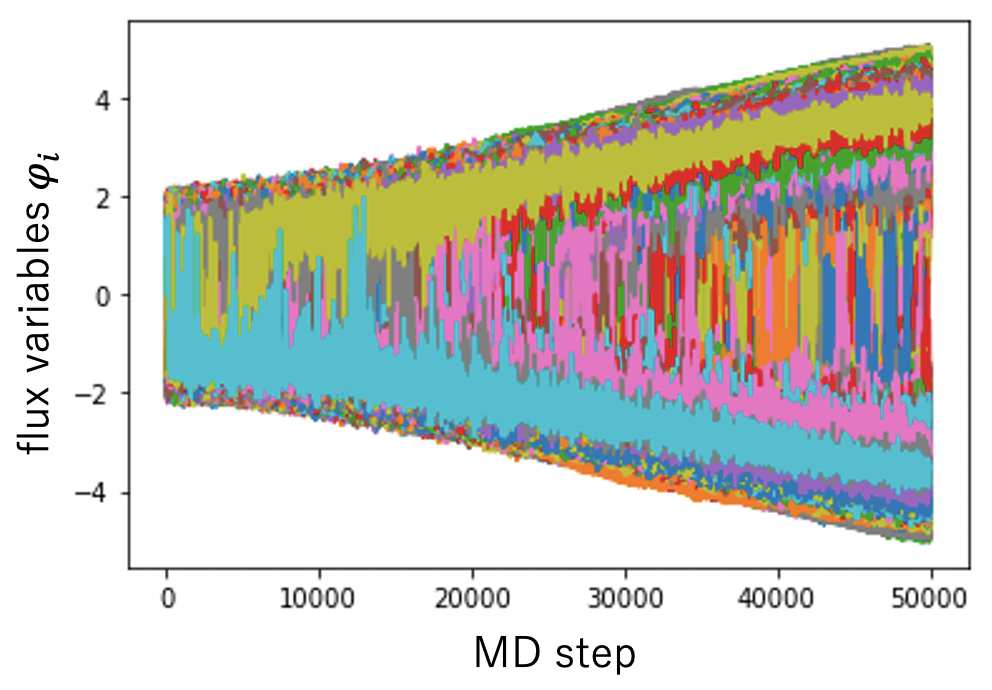}
\caption{Trajectories of variables $\{\varphi_i\}_{i=1}^{N}$ for a typical Ising spin-glass model with $N=10,000$ and 50,000 MD steps.}
\label{fig:flux_orbit}
\end{figure}

In Fig. \ref{fig:flux_orbit}, the trajectories of all flux variables are color-coded. 
By observing each trajectory, we can see that initially, they oscillate around $\varphi_i=0$. 
As time evolution progresses, some trajectories converge to small oscillations and transition in either positive or negative directions, while others continue to oscillate around $\varphi_i=0$. 
Therefore, when $\mathcal{H}_{\rm{MD}}$ is time-evolved, we can distinguish two tendencies in each trajectory: one group settles into positive or negative states, while the other group continues to oscillate around $\varphi_i=0$. 
In prior research, the former type of trajectories is referred to as "frozen variables" and the latter as "ambivalent variables."

When $\mathcal{H}_{\rm MD}$ in Eq. \eqref{HMD} is time-evolved using MD, we observe these two tendencies in the flux variable trajectories, with the frozen variables converging to positive or negative states and the ambivalent variables oscillating around zero.

The frozen variables can be considered to have determined their sign, which can be associated with the spin values. 
On the other hand, the ambivalent variables are still oscillating around zero, implying that their spin orientation has not been fixed yet. 
Therefore, prior research proposed a hybrid QA method. 
We refer to this method as MD+QA. 
In MD+QA, the frozen variables' signs are fixed as the spin values, while the ambivalent variables are solved using QA.

To properly distinguish between ambivalent and frozen variables, it is necessary to quantitatively characterize the tendencies of "positive or negative convergence" and "continued oscillation around $\varphi_i = 0$" in the trajectories and determine the variables as ambivalent or frozen. 
Therefore, in the previous research, the time average of the flux variables is calculated using the following equation:
\begin{equation}
\overline{\varphi}_{i}\equiv \frac{1}{\delta}\int_{1-\delta}^{1}d \tau \varphi_i(\tau).
\label{time average}
\end{equation}
The interval $\delta$ is chosen to be sufficiently large but much smaller than $1$, specifically $\delta \tau$ (the reciprocal of the total number of steps). 
Flux variables exhibiting continued oscillation around zero have both positive and negative values in the interval $\delta$. On the other hand, variables showing positive or negative convergence converge take values of only one sign or the other in the interval $\delta$.

As a result, the absolute value of $\overline{\varphi}_{i}$ for these variables is smaller than those that converge to positive or negative values. By sorting the values of $|\overline{\varphi}_{i}|$, we can distinguish ambivalent variables with smaller $|\overline{\varphi}_{i}|$ from frozen variables with larger $|\overline{\varphi}_{i}|$. 
Concretely, we sort the values as $|\overline{\varphi}_{1'}|\le|\overline{\varphi}_{2'}|\le \cdots \le |\overline{\varphi}_{n'}| \le \cdots \le |\overline{\varphi}_{N'}|$ (where $i'$ denotes the index after sorting). 
We can then consider $\{\varphi_{i'}\} \ (i'=1,\cdots,n)$ as ambivalent variables and $\{\varphi_{i'}\} \ (i'=n+1,\cdots,N)$ as frozen variables. The integer value $n\ (<N)$ that separates ambivalent and frozen variables can be preselected with a suitable value.

With the classification of flux variables into frozen and ambivalent variables, the frozen variables can be handled by simply extracting their signs as $s_{k'}={\rm sgn}(\overline{\varphi}_{k'})$, where ${\rm sgn}(x)$ is the sign function that returns $+1$ for $x > 0$ and $-1$ for $x \leq 0$. 
By fixing the signs in this way, the values of the frozen variables are determined and can be treated as spin orientations.

On the other hand, for the ambivalent variables, since their values are not yet determined, we proceed to solve the following Ising model using QA:
\begin{equation}
\begin{split}
&\mathcal{H'}_{\rm{Ising}}(s) = \frac{1}{2}\sum_{i'\ne j'}^{n}J_{i'j'}^{\rm{eff}}s_{i'}s_{j'}+\sum_{i'=1}^{n}h_{i'}^{\rm{eff}}s_{i'}  \\
    &\quad \equiv \mathcal{H}_{\rm{Ising}}(s|s_{k'=n+1,\dots,N} : \rm{frozen})-(\rm{const.}). 
\label{Hamiltonian QA}
\end{split}
\end{equation}
Here $J_{i'j'}^{\rm{eff}}$ and $h_{i'}^{\rm{eff}}$ are determined as follows by substituting the values of the frozen spins ${s_{i'}}\ (i'=n+1,\cdots, N)$:
\begin{equation}
J_{i'j'}^{\rm{eff}}=J_{i'j'}, \  h_{i'}^{\rm{eff}}=h_{i'}+\sum_{k'=n+1}^{N}J_{i'k'}s_{k'},\ (i',j'=1,2,\dots,n). 
\end{equation}
These effective interactions and fields take into account the contribution from the frozen spins and allow us to reformulate the Ising model for the ambivalent variables based on their interactions with the already fixed frozen spins. 
By solving this modified Ising model using QA, the values of the ambivalent variables can be determined and thus complete the hybrid QA approach.

The MD+QA method effectively complements the large-scale performance of QA by reducing the number of variables to be solved using QA through the preprocessing MD. 
However, there exists ambiguity in determining which variables are considered unstable based on their trajectories. Moreover, there are challenges regarding the accuracy of the variables fixed as frozen variables. Since frozen variables are not re-optimized by QA, their accuracy is crucial in this approach.

To address these issues, we propose a new hybrid approach that revises the continuous optimization method used in preprocessing and changes the method of extracting variables to be solved with QA. 
The following section introduces a typical combinatorial optimization problem to test our hybrid method.

\section{Problem Setting}
This section introduces the combinatorial optimization problem addressed in this study. 
We focus on the minimum vertex cover problem (min-VC) as a typical combinatorial optimization problem related to graphs. 
The minimum vertex cover problem aims to find a vertex cover in a graph $G=(V, E)$, where each edge connects two vertices, and at least one of the connected vertices must be covered by the selected vertices. 
The objective is to minimize the number of vertices in the cover.
The formulation of min-VC is as follows:
\begin{alignat}{3}
 & \text{minimize} & \sum_{i=1}^{N} x_{i},& \\
 & \text{subject to} \quad& x_i + x_j &\geq 1, &\forall (i,j) &\in E, \\
                 && x_{i}& \in \{0,1\},\quad &\forall i &\in V.
\end{alignat}
Here, the variable $x_i$ takes the value of $1$ when a vertex $i$ is selected and $0$ otherwise. 
The inequality constraints correspond to the requirement that "at least one of the vertices connected by each edge must be covered."

Furthermore, min-VC can be formulated as a quadratic unconstrained binary optimization (QUBO) problem as follows:
\begin{equation}
\text{minimize} \quad \sum_{i=1}^{N} x_{i} + \lambda \sum_{(i,j)\in E}(1-x_i)(1-x_j).
\label{QUBO of min-VC}
\end{equation}
This formulation relaxes the inequality constraints in the original equation and introduces the coefficient $\lambda$ to add to the objective function.

Next, we explain the properties of the min-VC problem. 
When applying the LP relaxation, the variables converge to $\{0, 1/2, 1\} $ \cite{nemhauser1974properties}. 
We introduce linear programming (LP) relaxation, which relaxes integer or binary variables into continuous variables. Here LP relaxation problem formulation for min-VC is expressed as follows:
\begin{alignat}{3}
\label{LP of min-VC}
 & \text{minimize} & \sum_{i=1}^{N} x_{i},& \\
 & \text{subject to} \quad& x_i + x_j &\geq 1, &\forall (i,j) &\in E, \\
                 && 0 \le x_{i} &\le 1,\quad &\forall i &\in V.
\end{alignat}
In the case of min-VC, by relaxing the binary variables $x_i \in \{0, 1\}$ to $0 \leq x_i \leq 1$, it becomes evident that the formulation becomes an LP.
Therefore, min-VC is a typical combinatorial optimization problem where LP relaxation can be applied.

In this study, we consider random graph instances of the min-VC problem. 
We consider connecting $N$ distinct pairs of nodes (totaling $N(N-1)/2$ pairs) with edges with a probability $p$.
The graph created in this way is called an Erd\"{o}s-R\'{e}nyi graph.

Here, a parameter $c = p \cdot N$ is introduced, and statistical-mechanics analysis\cite{Takabe_2016} has shown that the number of variables converging to $1/2$ rapidly increases at the critical value of $c=e=2.71$. 
In other words, when $c$ is smaller than $e$ (indicating a sparser graph), all variables converge to $\{0,1\}$ through LP relaxation. 
In this case, the LP relaxation provides the exact solution to min-VC. 
On the other hand, when $c$ is larger than $e$ (indicating a denser graph), all variables converge to $\{0,1/2,1\} $ through LP relaxation. 
Consequently, in this case, min-VC cannot be solved exactly using LP relaxation.

\begin{figure}[tbh]
\centering
\includegraphics[scale=0.5]{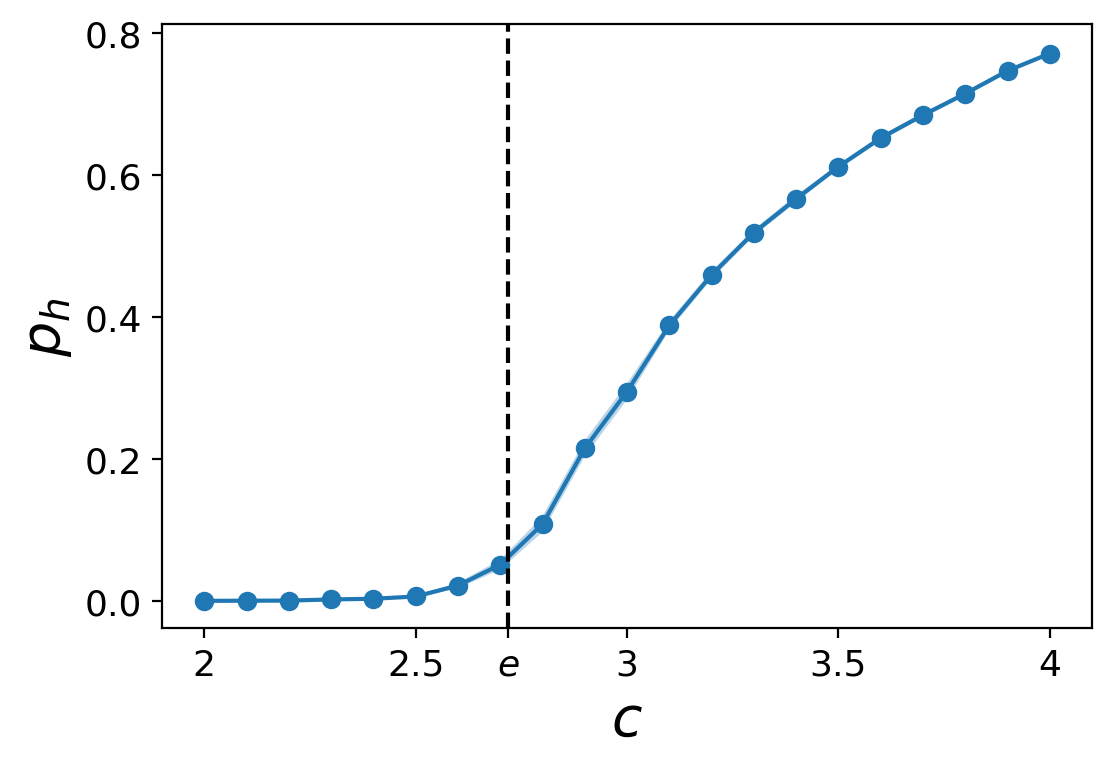}
\caption{$c$-dependency of $p_h$ in min-VC($N=2000$).}
\label{fig:half_ratio}
\end{figure}

Figure \ref{fig:half_ratio} illustrates the dependency of the rate of variables converging to 1/2 on the parameter $c=p \cdot N$.

The plot shows the results of solving min-VC through LP relaxation for Erd\"{o}s-R\'{e}nyi graphs with $N=2000$. 
The vertical axis, $p_h$, represents the proportion of variables that converge to $1/2$. 
The LP relaxation problems were solved using the Gurobi Optimizer (version 10.0.0). 
From the graph, it can be observed that, for instance, when $c=3$, approximately $30\%$ of the variables converge to $1/2$, and when $c=4$, around $70-80\%$ of the variables converge to $1/2$.

In the next section, we describe our hybrid algorithm of LP relaxation and QA. 
In this method, we apply the properties of the LP relaxation of min-VC described above.
\section{Method}
Our developed hybrid approach combining LP relaxation and QA is called LP+QA. 
In this method, we replace the MD part of MD+QA with LP relaxation. 
The specific procedure for LP+QA is as follows:
\begin{enumerate}
   \item Solve min-VC using LP relaxation.
   \item Extract variables that converge to $1/2$ from the obtained solution.
   \item Re-optimize the extracted variables using QA.
\end{enumerate}

In Step 1, we solve the LP relaxation of the min-VC problem given by Eqs. (13)-(15) using algorithms such as the simplex or interior point methods. 
These algorithms efficiently find solutions for large min-VC instances, and the solutions obtained are deterministic, allowing for exact results for the LP relaxation problem. 
However, in the min-VC problem, several variables may converge to $1/2$ depending on the parameter $c$. 

In Step 2, we extract those solutions where variables converge to $1/2$. 

In Step 3, we solve only those extracted variables using QA, following the same approach as in MD+QA. 
In other words, we design a partial Ising model in which only the variables converging to $\{0,1\}$ are assigned values based on Eqs. (8) and (9), leaving the variables converging to $1/2$ as decision variables.

Moreover, the QA part in Step 3 can be replaced with other heuristics, such as simulated annealing (SA) \cite{Kirkpatrick1983}.

This approach utilizes the property that when solving min-VC using LP relaxation, several variables converge to $1/2$. 
In MD+QA, the method involved continuous optimization using MD, then classifying variables into ambivalent and frozen variables and re-solving ambivalent variables. 
In our approach, we precisely extract variables that converge to half-integer values and re-optimize them, making removing "difficult" variables more straightforward and explicit. 
However, since the convergence property of the LP relaxation solution to $\{0, 1/2, 1\}$ is specific to the min-VC, it should be noted that our method is at least limited to min-VC.

\section{Result}
To assess the performance of our method, we perform numerical experiments on the min-VC problem for Erd\"{o}s-R\'{e}nyi graphs with the parameter $c=p\cdot N$.

First, we compare LP+QA, LP+SA, MD+QA, and MD+SA. Here MD+SA ( or LP+SA ) replaces the QA portion of MD+QA ( or LP+QA ) with SA. The conditions for each solver are as follows:

\begin{itemize}
      \item Gurobi Optimizer (version: 10.0.0) is used to solve the LP relaxation problem. 
      \item D-Wave Advantage 6.2 is used for QA, with an annealing time of $20\ {\rm \mu s}$ and obtaining $100$ samples.
      \item We use Openjij as the SA package with $100$ samples and default parameters.
      \item  As the MD parameters of Eq. \eqref{schedule_alpha} and Eq. \eqref{schedule_beta}, we set $(\alpha_f, \rho_1, \rho_2)=(0.008,4,3) $ and $(\beta_f, \kappa_1, \kappa_2)=(0.12, 0.005, 1) $. And also the parameter $M$ for the potential term $V(\varphi)=\varphi^{M}$ in Eq. \eqref{HMD} is $M=6$. These parameters were used in the numerical experiments of previous research \cite{Irie_HQA_2021}. The total MD steps is $500,000$.
\end{itemize}

The penalty coefficients, $\lambda$, for the QUBO formulation of min-VC are set as $(\lambda_{\rm MD}, \lambda_{\rm QA}, \lambda_{\rm SA}) = (2, 1, 1)$. 
These values were tuned for each solver.
Under these conditions, we solved $100$ instances of the min-VC problem for Erd\"{o}s-R\'{e}nyi graphs with $N=1000, c=3$ using the four solvers mentioned above. 

In both MD+QA (or MD+SA) and LP+QA (or LP+SA), we ensure that the number of variables solved with SA or QA is the same. 
As we have $c=3$ in this case, according to Fig. \ref{fig:half_ratio}, the number of variables converging to $1/2$ is around $30\%$. 
Hence, we solve approximately $300$ variables using SA or QA. However, it is important to note that the number of variables converging to $1/2$ varies for each random graph instance. 
The number of variables to be re-optimized may differ from one instance to another.

\begin{figure}[tbh]
\centering
\includegraphics[scale=0.5]{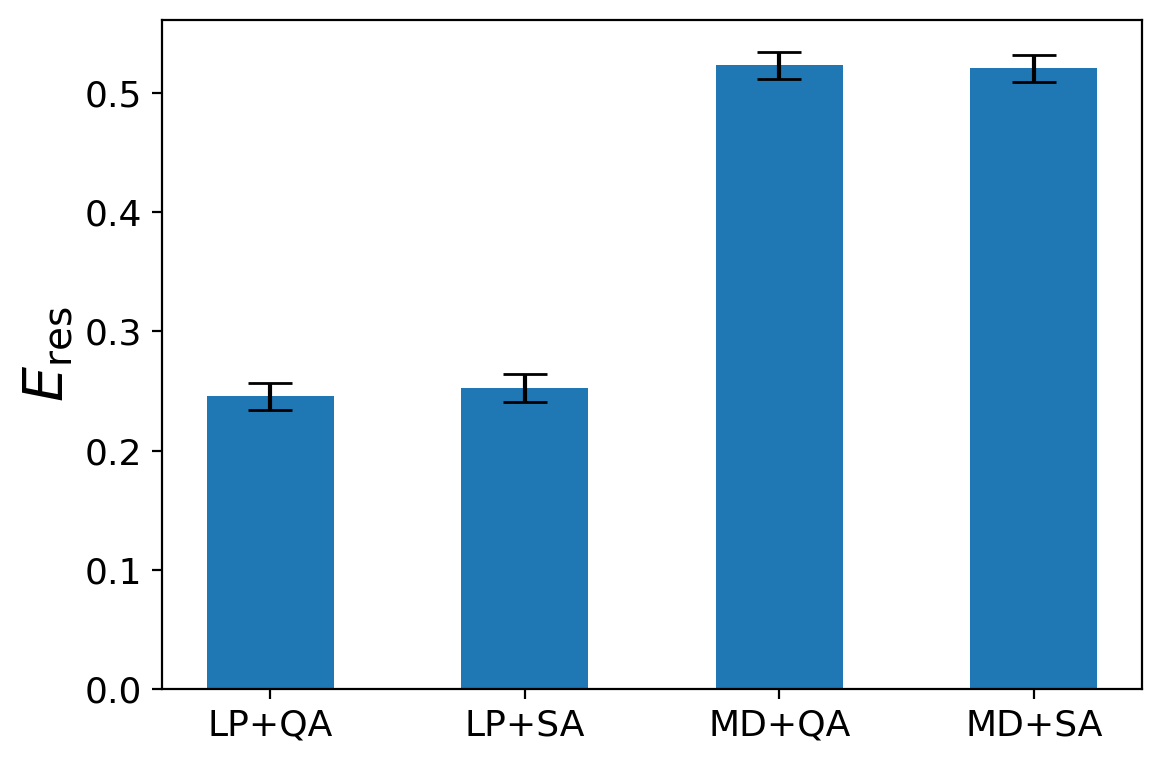}
\caption{The comparison results for the min-VC problem with LP+QA, LP+SA, MD+QA, and MD+SA in min-VC($N=1000, c=3$).}
\label{fig:4method_compare}
\end{figure}

Figure \ref{fig:4method_compare} shows the residual energy of LP+QA, LP+SA, MD+QA, and MD+SA. 
Here the residual energy $R_{\rm res}$ is defined by $E_{\rm res} = (E - E_{\rm opt})/{|E_{\rm opt}|}$, where $E$ is the value of the objective function obtained from each solver and $E_{\rm opt}$ is the exact value. 
We obtain the exact solution by Gurobi optimizer. 

First, comparing the QA-hybrid approach, MD+QA, with LP+QA, we can observe that the proposed LP+QA method achieves better accuracy than the existing MD+QA. 
Therefore, our approach proves to be effective in solving large-scale problems using the D-Wave quantum annealer. 
In addition, there was almost no difference in accuracy between MD+QA and MD+SA and between LP+QA and LP+SA.

In the above experiments, we ensured that the number of variables to be solved using QA or SA was the same for MD- and LP-based methods. 
This was because, in the LP-based approach, we cannot freely choose the number of variables to be re-optimized. 
On the other hand, in the MD-based approach, we can divide the variables into ambivalent and frozen ones using the parameter $n$, allowing us to select the number of variables to be solved as desired. 
However, in the comparison, we aligned the number of variables to be solved with the LP-based approach, which might lead to a bias in favor of the LP-based method. 
Therefore, in the next experiment, we investigate the $n$-dependency of the accuracy particularly for the MD-based approach. 
We also show how the accuracy difference between the LP-based and MD-based methods changes as $n$ varies.

In the above experiments, we observed almost no difference in accuracy between using SA and QA for both LP-based and MD-based methods. 
Therefore, we will focus on performing experiments especially with SA in this experiment.

\begin{figure}[tbh]
\centering
\includegraphics[scale=0.5]{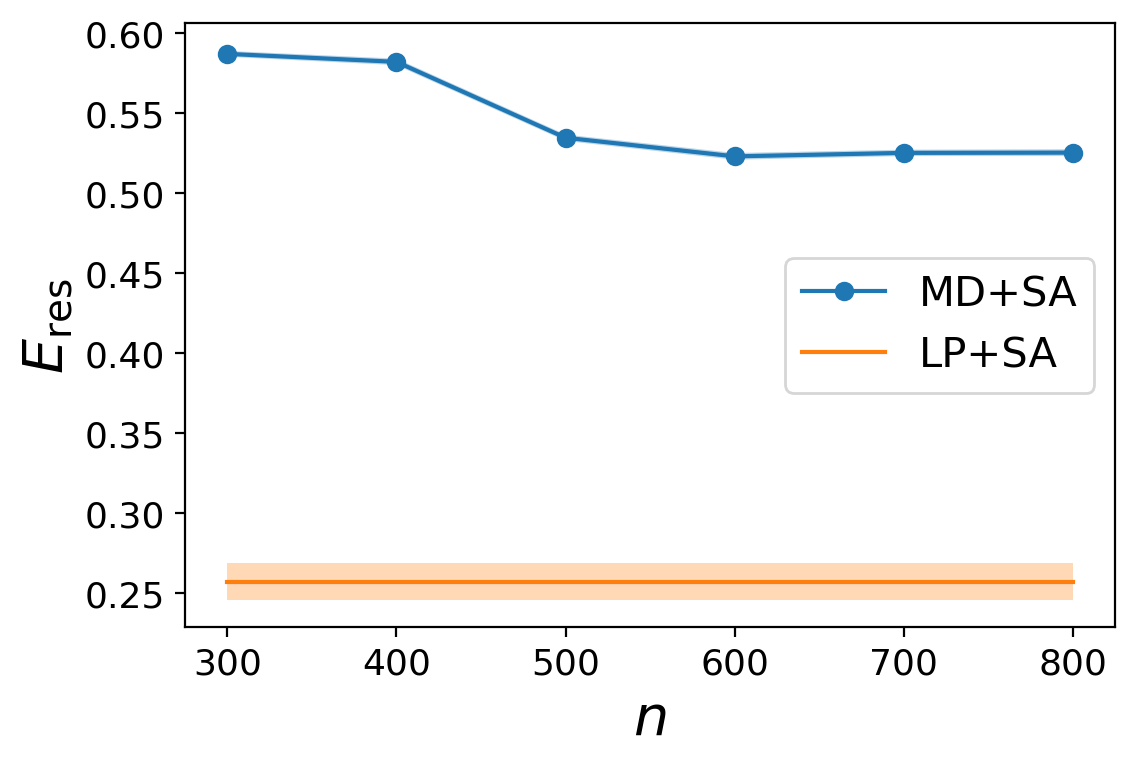}
\caption{$n$-dependency of the residual energy of MD+SA in min-VC($N=1000, c=3$). $n$ is the number of variables solved by SA in MD+SA.}
\label{fig:MDSAvsLPSA}
\end{figure}

Figure \ref{fig:MDSAvsLPSA} shows the $n$-dependency of residual energy $R_{\rm res}$ of MD+SA, and LP+SA in min-VC($N=1000, c=3$). 

In Fig. \ref{fig:MDSAvsLPSA}, we show the $n$-dependency of the residual energy of MD+SA with blue plots and the residual energy of LP+SA with an orange line. Since $n$ is a parameter specific to the MD+SA method, the results for LP+SA form a parallel line along the horizontal axis. As the value of $n$ increases, the accuracy of MD+SA gradually improves. However, it reaches a plateau around $n=600$. 
It suggests that, regardless of the MD preprocessing, the performance of MD+SA becomes nearly equivalent to SA-only. 
Furthermore, this result implies that even with increasing values of $n$ in MD+SA, it does not achieve the same level of accuracy as LP+SA. 
Therefore, our proposed method demonstrates superiority over the existing approaches.

Next, we examine whether our SA-hybrid approach exhibits advantages over solving the min-VC with SA only. 
This experiment compares the computational time and objective function values between LP+SA and SA-only. 

SA-only refers to the method of solving the QUBO of min-VC, as represented by Eq. \eqref{QUBO of min-VC}, without the preprocessing of LP relaxation. 
In other words, in this experiment, we investigate whether our approach as an SA-hybrid method demonstrates advantages in terms of accuracy or computational speed compared to the SA-only approach.

We obtain the computational time for solving the LP relaxation problem from the module that provides computational time included in the Gurobi package. Similarly, the computational time for SA is obtained from the module included in the Openjij package. 
In addition, for LP+SA, the computational time is simply the sum of the time to solve the LP relaxation problem obtained from the Gurobi module and the SA computational time obtained from Openjij.

\begin{figure}[tbh]
\centering
\includegraphics[scale=0.5]{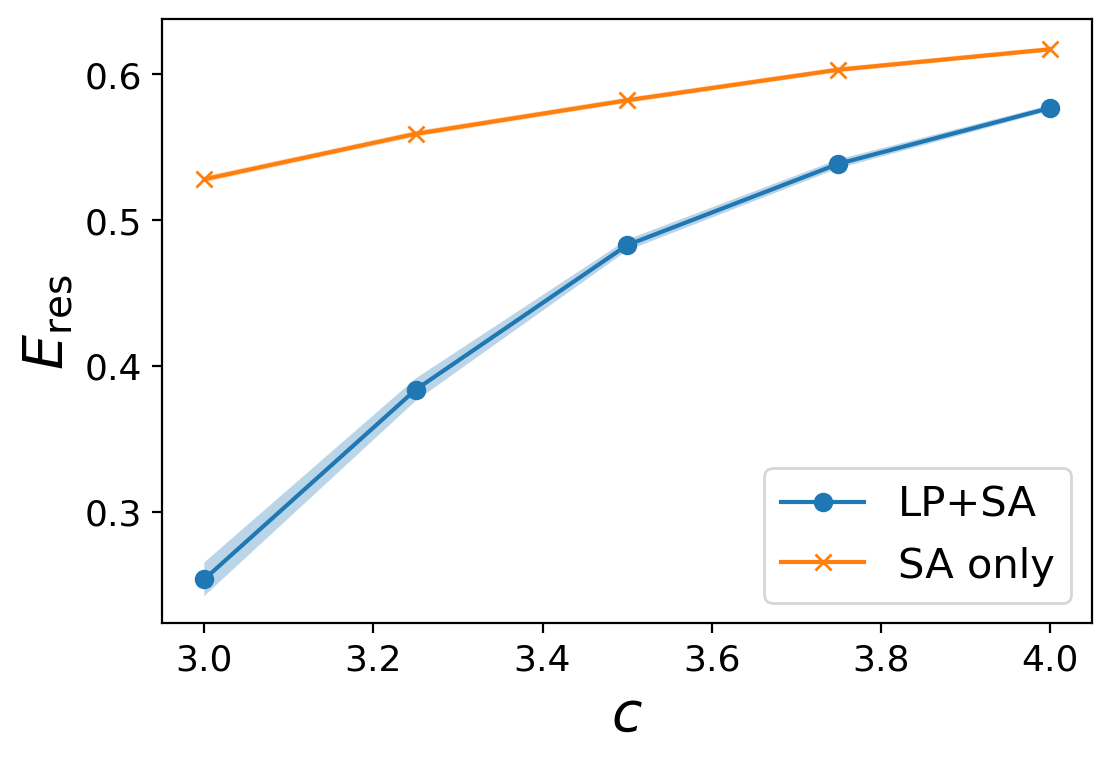}
\caption{$c$-dependency of the residual energy of LP+SA and SA-only in min-VC($N=1000$).}
\label{fig:SA_vs_LPSA_c_dpd}
\end{figure}

Figure  \ref{fig:SA_vs_LPSA_c_dpd} shows the $c$-dependency of the residual energy $R_{\rm res}$ of LP+SA and SA-only. 

From Fig. \ref{fig:SA_vs_LPSA_c_dpd}, it can be observed that LP+SA achieves higher accuracy than SA-only for all values of $c$. 
However, as $c$ increases, the solutions obtained by LP+SA approach the accuracy of SA-only solutions, as indicated in the figure. This behavior is attributed to the decrease in the number of variables that can be fixed through the LP relaxation preprocessing, as shown in Fig. \ref{fig:half_ratio}.

\begin{figure}[tbh]
\centering
\includegraphics[scale=0.55]{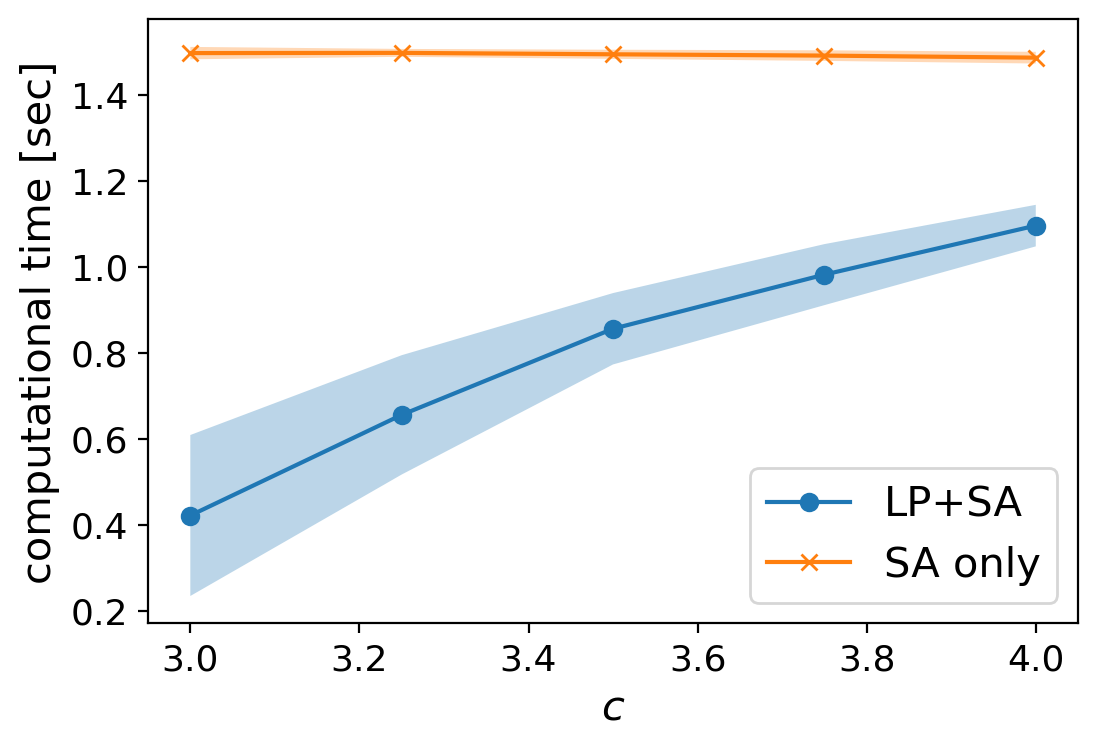}
\caption{$c-$dependency of computational time of LP+SA and SA-only in min-VC($N=1000$).}
\label{fig:time_c_dpd}
\end{figure}

Figure \ref{fig:time_c_dpd} shows the $c$-dependency of computational time. 
This computational time is measured as each solver's time to obtain the results shown in Fig. \ref{fig:SA_vs_LPSA_c_dpd}.

From Fig. \ref{fig:time_c_dpd}, it can be observed that SA-only exhibits nearly constant computational time across varying values of $c$ due to the fixed problem size $N=1000$. 
On the other hand, LP+SA shows an increasing computational time with increasing $c$. 
This is because as $c$ increases, the number of variables converging to $1/2$ increases. 
This implies that the reduction in the number of variables achieved by LP preprocessing decreases, requiring more variables to be solved using SA. However, since the Gurobi Optimizer efficiently solves the LP relaxation problems in the preprocessing step, LP+SA remains faster than SA-only. 
In addition, as shown in Fig. \ref{fig:SA_vs_LPSA_c_dpd}, LP+SA exhibits better accuracy.

Evidently, around $c=3$, LP+SA outperforms SA-only regarding speed and accuracy based on these results. 
Therefore, our method proves to be effective as an improvement to the SA-hybrid approach.

\section{Discussion}
We performed experiments using min-VC to compare our method with MD+QA. 
As a result, our approach outperformed MD+QA in terms of accuracy. 
We believe that the difference in the accuracy of their respective preprocessing steps could contribute to this outcome. 
In particular, as mentioned in previous studies, MD approximately estimates the ground state of the Ising model. 
There are errors in the frozen variables, i.e., fixed variables, without reevaluating them using QA. 
On the other hand, when solving with min-VC using LP relaxation, the convergence of variables to $\{0,1\}$ is more accurate. 
Therefore, we expect our method's accuracy as an optimizer is higher than MD.

Furthermore, when comparing LP+QA and LP+SA, there was almost no difference in accuracy. The min-VC problem with $c=3$ is a sparse graph problem, making it relatively easy to embed into the topology of the D-Wave machine. 
Therefore, under these conditions, LP+QA can be utilized as a method to obtain solutions of equivalent accuracy to LP+SA faster.

Our approach has certain limitations. Since we utilize LP relaxation, it only applies to combinatorial optimization problems that can be solved through it. 
In particular, problems with quadratic terms in the objective function, such as the max-cut problem, cannot be solved using LP relaxation.

In this study, we dealt with min-VC problems where the variables neatly converged to $\{0,1/2,1\}$ through LP relaxation, making it straightforward to determine which variables needed to be re-optimized using QA or SA. However, for other combinatorial optimization problems, non-integer values other than $1/2$ may arise during LP relaxation. 
In such cases, careful consideration is required to select the variables to be re-optimized.

One possible method to handle a broader class of problems using the QA-hybrid approach is to apply semi-definite programming (SDP) relaxation as an alternative to LP relaxation. 
It converts the original problem into a semi-definite programming problem by relaxing it to continuous variables, making it suitable for handling quadratic programming problems like the max-cut problem. 
However, SDP relaxation may also converge to non-integer values other than $1/2$, necessitating consideration of selecting the variables for re-optimization.

In addition, the simulated bifurcation (SB) algorithm \cite{goto2019combinatorial} is a method that solves combinatorial optimization problems by continuously relaxing them and subsequently determining discrete variables. 
Similarly, the challenges of choosing re-optimization variables apply when converting the continuous relaxation solutions to discrete variables using SB. 
Therefore, comparing our method with SB also presents an interesting direction for future exploration.

Another relevant method closely related to our approach is the branch and bound method, which is widely used to find exact solutions for combinatorial optimization problems. 
In this approach, LP relaxation problems are solved, and based on the results, branching is performed by assigning $0$ or $1$ to the variables that converged to $1/2$, leading to precise solutions. 
In contrast, our method employs heuristics like QA or SA to obtain approximate solutions more roughly but rapidly. 
With increasing problem size, our method is expected to exhibit a computational advantage over the branch and bound method, as the latter may require substantial time to reach the precision achieved quickly by our approach.

{\it Acknowledgement.}
The authors thank the fruitful discussions with Y. Nishikawa and J. Takahashi.
Our study receives financial support from the MEXT-Quantum Leap Flagship Program Grant No. JPMXS0120352009, as well as Public\verb|\|Private R\&D Investment Strategic Expansion PrograM (PRISM) and programs for Bridging the gap between R\&D and the IDeal society (society 5.0) and Generating Economic and social value (BRIDGE) from Cabinet Office.

\bibliographystyle{jpsj}
\bibliography{main}
\end{document}